\documentclass[prl,twocolumn,showpacs,superscriptaddress]{revtex4}
\usepackage{graphicx}

\begin{document}

\title{Fluctuation analysis of mechanochemical coupling depending on the type of bio-molecular motor}
\date{\today}
\author{Masatoshi Nishikawa}
\affiliation{
Graduate School of Frontier Biosciences, Osaka University, 1-3 Yamadaoka, Suita Osaka 565-0871, Japan
}
\author{Hiroaki Takagi}
\affiliation{
Department of Physics, Nara Medical University, 840 Shijo-cho, Kashihara Nara  634-8521, Japan
}
\author{Atsuko H. Iwane}
\affiliation{
Graduate School of Frontier Biosciences, Osaka University, 1-3 Yamadaoka, Suita Osaka 565-0871, Japan
}
\author{Toshio Yanagida}
\affiliation{
Graduate School of Frontier Biosciences, Osaka University, 1-3 Yamadaoka, Suita Osaka 565-0871, Japan
}

\begin{abstract}
Mechanochemical coupling was studied for two different types of myosin motors in cells: myosin V, which carries cargo over long distances by as a single molecule; and myosin II, which generates a contracting force in cooperation with other myosin II molecules. Both mean and variance of myosin V velocity at various [ATP] obeyed Michaelis-Menten mechanics, consistent with tight mechanochemical coupling. Myosin II, working in an ensemble, however, was explained by a loose coupling mechanism, generating variable step sizes depending on the ATP concentration and realizing a much larger step (200 nm) per ATP hydrolysis than myosin V through its cooperative nature at zero load. These different mechanics are ideal for the respective myosin's physiological functions.
\end{abstract}


\maketitle

Molecular motors play essential roles in various physiological functions, such as muscle contraction, molecular transport, cell motility and cell division through displacements powered by the chemical energy of ATP hydrolysis.
The primary goal of molecular motor biophysical studies is to understand how chemical energy is converted into mechanical movement, i.e. mechanochemical coupling.
Single molecule techniques are a powerful tool to investigate chemical reactions and mechanical movements by molecular motors because they can directly observe the elemental reaction process that is undetectable in ensemble average measurements \cite{Funatsu95,Svoboda93}.
This is especially true for motors like myosin V, which function as a single molecule or in cooperation with a small number of other myosin V molecules \cite{Mehta99}.
By combining single molecule studies with \textit{in vitro} kinetics studies, myosin V is thought to produce regular 36 nm steps coupled to one ATP hydrolysis \cite{Rief00}. This means that the mechanochemical coupling of myosin V is a one-to-one relationship, i.e. "tight coupling", although no direct measurement has shown this relationship.
In contrast, little is known about the mechanochemical coupling of ensemble functioning motors, such as myosin II in muscle. Because of its poor stepping ability, mechanical steps are easily buried in thermal noise. This makes direct detection of individual mechanical events difficult \cite{Kitamura99}, leaving most to discuss only the average displacement \cite{Molloy95}. 
Furthermore, because \textit{in vivo} myosin II work in ensemble, single molecule characteristics may not accurately reflect the mechanochemical coupling of myosin II.
In fact, multi-molecule systems show various characteristics unpredictable from single molecule studies \cite{Fenn24, Yasuda96}. 
Therefore, the mechanochemical coupling of myosin II is still poorly understood.

\begin{figure}[htbp]
\includegraphics[width=8cm]{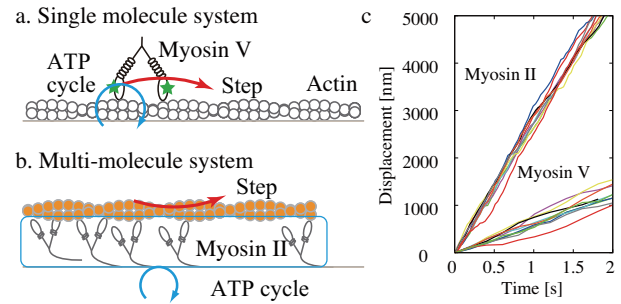}
\caption
{Schematic of the experiments and typical traces of myosin motility.
\textit{a}, Single molecule motility assay of myosin V.
The single molecule movement of myosin V along the actin filament was directly measured.
Myosin molecules were genetically fused GFP (Green Fluorescent Protein) for observation 
(green star). Fluorescence from myosin in solution is not imaged because of its fast diffusion whereas molecules moving along actin filaments are imaged as moving bright spots. The experiments were performed at 25 $^\circ$C.
\textit{b}, \textit{In vitro} actin gliding assay. The movement of an individual actin filament
driven by a myosin II ensemble was measured. 
Actin filaments were attached to the fluorescent dye tetramethyl - rhodamine phalloidin for observation
(colored in orange). Myosin molecules were not labeled and not observed. The experiments were performed at 28 $^\circ$C.
\textit{c}, Typical displacement time series obtained from the motility assays.
ATP concentration, [ATP], was 100 $\mu$M in both assays.}
\label{fig1}
\end{figure}

In order to examine mechanochemical coupling of both kinds of molecular motors, in this report, we investigated the characteristics of fluctuation in myosin V and myosin II motility and discuss their mechanochemical coupling consistently.
Since these fluctuations contain the underlying molecular process, we can extract pertinent information of mechanochemical coupling from fluctuation analysis, as Schnitzer et al. did to show one kinesin step couples to a single ATP turnover \cite{Schnitzer97}.

In myosin V experiments, we directly observed single myosin molecule movement along actin filaments \cite{M06}. Fluorescence imaging was performed by TIRF microscopy \cite{Tokunaga97}. Actin filaments were adsorbed onto a cover slip. Freely floating fluorescently dyed myosins in solution bound to actin filaments commencing movement. The movement was tracked by fitting the fluorescent spots with a two-dimensional Gaussian distribution function.
On the other hand, in myosin II experiments, we performed the \textit{in vitro} actin gliding assay \cite{Uyeda90} (Fig. \ref{fig1}) and measured the sliding velocity between the actin filament and myosin II molecules. Myosin II molecules were adsorbed onto the substrate. Introducing actin filaments with fluorescent dye and ATP resulted in a sliding movement between the myosin molecules and the actin filaments. The trajectories of the actin movement were obtained by tracking the center of the brightness of the actin fluorescence. For the convenience of the analysis, we chose the length of the actin filament to be less than 600 nm.
In order to obtain velocity statistics, we converted two dimensional trajectories of both experiments into a displacement time series and calculated the Mean Square Displacement (MSD). 
We fitted the MSD to MSD$(\Delta t) = \mu_v^2 \Delta t ^2+ \sigma_v^2 \Delta t  + \xi$, 
where $\mu_v$ is the mean velocity, $\sigma_v^2$ is the variance of velocity per unit time, 
 $\xi$ is the measurement error, and $\Delta t$ is the sampling time.
We performed both experiments at various ATP concentrations, calculated the MSDs for each [ATP], and identified the above parameter values. Here, we set $\Delta t$ = 200 ms for myosin V and 32 ms for myosin II.
For myosin V, MSDs were calculated from 100 traces, each of which contained more than 12 sampling points on average at [ATP] ranging from 2 to 1000 $\mu$M; for myosin II, MSDs were calculated from 30 traces, each of which contained more than 100 sampling points on average at all [ATP] conditions. In very low [ATP] conditions (200 nM to 1 $\mu$M), we calculated myosin V MSDs from 40 traces, each of which contained more than 30 sampling points on average at 
$\Delta t$ = 2 s.
  
The relationship between [ATP] and the mean velocity is shown in Fig. \ref{fig2} . This relationship is known as the Michaelis - Menten (MM) mechanism, which represents the catalytic activity of the enzyme.  We applied the MM mechanism to the actomyosin sliding movement :

\begin{figure}[htbp]
\includegraphics[width=8cm,clip]{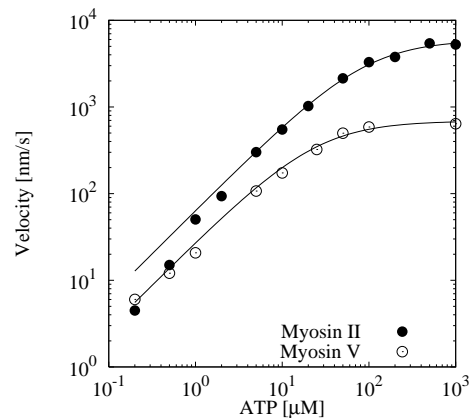}
\caption
{The dependency of the mean velocity on the ATP concentration in motility assays.
Solid lines show the fit of the Michaelis-Menten equation to the experimental data.
See text for details.}
\label{fig2}
\end{figure}

(Scheme 1)  \[A\cdot M + ATP \leftrightarrow A\cdot M\cdot ATP \rightarrow A\cdot M + Pr \]
, where M, A, and Pr denote myosin, actin, and the ATP hydrolysis products ADP and Pi, respectively. In this scheme, an ATP binds reversibly to the actomyosin complex. The actomyosin catalyzes ATP hydrolysis to produce a regular displacement, i.e. tight coupling.
Note that in the case of myosin II, a single actin filament interacting with multiple myosin II molecules follows Scheme 1, whereas in the case of myosin V, single myosin V molecule obeys MM kinetics. Although in the former case, the number of interacting myosin II molecules seems to affect the frequency of the chemical reaction, it was experimentally confirmed that the velocity of the single actin filament saturates and is independent of the number of interacting myosin molecules except for myosin densities much low than those used here \cite{Uyeda90,Harada90}.
In this scheme, the ATP dependence of the sliding velocity is described by $V=V_{max}[ATP]/(K_m+[ATP])$ , where $V_{max}$ is the maximal velocity at saturating [ATP] and $K_m$ is the [ATP] corresponding to the half maximal velocity of the sliding movement, called the Michaelis constant. Solid lines in Fig. \ref{fig2} show the fitted MM equation with  
$V_{max} = 689$ $nm/s$ and $K_m = 24.6$ $\mu$M for myosin V and $V_{max} = 5960$ $nm/s$ and $K_m = 93.0$ $\mu$M for myosin II, respectively.
These values are consistent with earlier studies \cite{Harada90, Baker04}.  This shows that regarding mean velocity, Scheme 1 holds for both kinds of myosins.

\begin{figure}[tbp]
\includegraphics[width=7.5cm,clip]{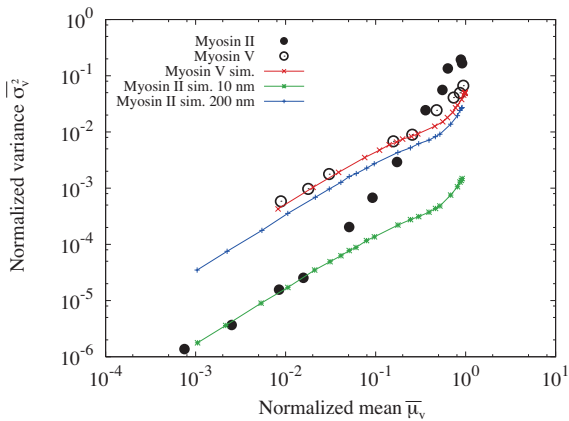}
\caption
{Relationship between $\mu_v$ and $\sigma_v^2$. To show the dependence of the velocity variance per unit time on $\mu_v$, we normalized the following variables as follows :
 $\bar{\mu_v}$ = $\mu_v$ / $V_{max}$, $\bar{\sigma_v^2}$ = $\sigma_v$ / $V_{max}^2$.
 Lines show the results of stochastic simulations of the MM mechanism. Parameters used in the simulations are as follows: $k_-$ = 18.5 $s^{-1}$ and $k_+$ = 0.750 $s^{-1}\mu M^{-1}$ for myosin V, $k_-$ = 596 $s^{-1}$ and $k_+$ = 6.40 $s^{-1}\mu M^{-1}$ for myosin II with 10 nm step, and $k_-$ = 30.0 $s^{-1}$ and $k_+$ = 0.320 $s^{-1}\mu M^{-1}$ for myosin II with 200 nm.
}
\label{fig3}
\end{figure}

Then, is this simple MM mechanism also valid for the description of the velocity fluctuation?
 In order to examine this point, we further investigated the relationship between $\mu_v$ and $\sigma_v^2$ (Fig. \ref{fig3}). 
In the case of myosin V, there is a linear relationship between $\mu_v$ and $\sigma_v^2$ (Open circle) except for the velocity region corresponding to the [ATP] around the Michaelis constant. 
To test the validity of Scheme 1, we performed stochastic simulations and compared these with the experimental results. 
Scheme 1 requires three rate constants even though experimentally only two parameters, $V_{max}$ and $K_m$, were obtained.
Therefore we simplified Scheme 1 without losing any relevant statistics:

(Scheme 2)  \[A\cdot M + ATP \leftrightarrow A\cdot M\cdot ATP\]
Since the left and right states in Scheme 1 are identical for the actomyosin complex, we can rewrite Scheme 1 as Scheme 2. Here we assumed a regular stepping motion is accompanied by a left directed reaction. Although the left directed arrow should include both ATP unbinding and ATP hydrolysis reaction, the former only reduces the stepping rate. 
This causes the reduction of the velocity at each [ATP], but does not affect the relationship between the mean and the variance of the velocity. Thus we can ignore the ATP unbinding reaction in this analysis.
The rate constant of the left directed reaction, $k_-$, was set to be $k_-$ =$V_{max}$ / step-size in order to obtain the experimental $V_{max}$. The rate of the right directed reaction, $k_+$, was calculated by $k_+$ = $k_-$ / $K_m$. We calculated the number of stepping events for a unit time interval in the stochastic simulation with the Gillespie algorithm \cite{Gillespie1977}. The step-size of myosin V was set at 36 nm, based on an earlier study \cite{Rief00}. 
The simulation results (red line in Fig. \ref{fig3}) show good agreement with the experimental data. This confirms that myosin V is a tight coupling motor, consistent with earlier studies.

The resulting relationship between $\mu_v$ and $\sigma_v^2$ is explained as follows : in very low [ATP] conditions, ATP binding is the single rate limiting step in the reaction sequence. Thus, the number of occurring chemical reactions in a given time interval can be described by a Poisson distribution, in which the linear relationship between $\mu_v$ and $\sigma_v^2$ holds.
On the other hand, when [ATP] is near the Michaelis constant, the frequency of ATP binding and the hydrolysis reaction are comparable. Here, the fluctuation of each reaction cancels, and the fluctuation decreases.  In higher [ATP] conditions, however, ATP hydrolysis becomes a single rate limiting step and $\sigma_v^2$ again becomes proportional to $\mu_v$. (We also confirmed the change of the number of the rate limiting steps through the randomness parameter analysis \cite{Schnitzer97}).

In the case of myosin II, the relationship between $\mu_v$ and $\sigma_v^2$ is qualitatively different from that of myosin V. $\sigma_v^2$ is proportional to $\mu_v$ in the lower velocity range, while $\sigma_v$ is proportional to $\mu_v$ in the intermediate and higher velocity range. 
The crossover of the relationship occurs at the velocity corresponding to [ATP] = 2 $\mu$M, which is similar to the Michaelis constant of ATPase \cite{Harada90}. 
We also performed stochastic simulations for Scheme 2 for the myosin II data. 
Here, the movement of a single actin filament is assumed to be displaced 10 nm by individual myosin II molecules, as is expected from the tight coupling model \cite{Spudich94} (Fig. \ref{fig3} in green). 
The simulation results of scheme 2 showed a linear relationship between $\sigma_v^2$ and $\mu_v$ that could not reproduce the crossover or the larger fluctuations appearing in the experimental data. 
The change in the number of interacting myosin II molecules i.e. the number fluctuation cannot produce the crossover for the following reasons: the number fluctuation has a temporal correlation in the time scale of $L/V$, where $L$ is the length of the actin filament and $V$ is the velocity. This becomes smaller as the [ATP] increases. When considering the velocity fluctuation per unit time, the effect of the number fluctuation also becomes smaller as the [ATP] increases because it is averaged out by the
law of large numbers. Furthermore, to achieve the $V_{max}$ value with a 10 nm step, we have to set the stepping rate to 600 $s^{-1}$, which is large enough to average out the
fluctuation of reactions. 
Thus, it is necessary to have a displacement larger than 10 nm per one ATP hydrolysis to reduce the number of stepping events per unit time and realize the larger fluctuations (for example, a 200 nm step simulation is shown in Fig. \ref{fig3} in blue). 
This cannot be achieved in a tight coupling model because it is not physically plausible for the myosin molecule to generate a regular displacement larger than its own body ($\approx $20 nm). 
Therefore, to produce a step-size more than 20 nm, multiple stepping events are required to occur in a single ATP hydrolysis. 

\begin{figure}[tb]
\includegraphics[width=7cm,clip]{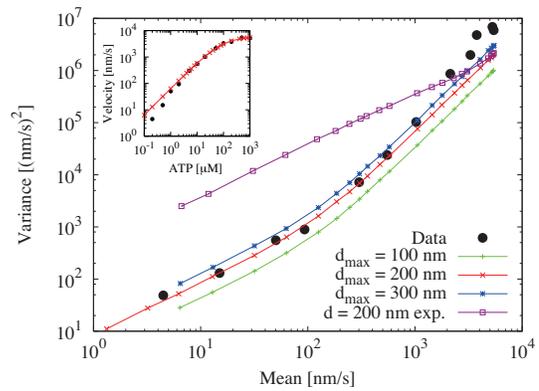}
\caption{
Comparison with our proposed model to the experimental data of myosin II. Symbols are the same as the myosin II data in Fig. \ref{fig3}. Solid lines show the simulated relationship of our model. See text for details.
\textit{Inset}, The dependence of the mean velocity on the [ATP]. Symbols are the same as myosin II data in Fig. \ref{fig2}. Red solid line is the stochastic simulation. The parameters are $d_{max}$ = 200 nm, $K_m^{ATP} = 2 \mu$M, $K_m^{mot} = 93 \mu$M, respectively.
}
\label{fig4}
\end{figure}

Employing a larger step-size can reproduce the large fluctuation as shown in Fig. \ref{fig3}. However, it still cannot reproduce the crossover relationship, because the stochastic property of Scheme 2 only holds the $\mu_v \propto \sigma_v^2$ relationship, as mentioned above. Therefore, we incorporated a variable step-size depending on the [ATP], taking into account the change in the relationship between $\mu_v$ and $\sigma_v^2$ around the Michaelis constant of the ATPase, 2 $\mu$M. Here we assume that the ATPase kinetics follows the Michaelis constant of 2 $\mu$M. This assumption leads to the relationship $V_{mot} = V_{ATPase} \times d$, where $d$ is the step-size per one ATP hydrolysis. Following from Scheme 2, $V_{mot} = V_{max}^{mot}[ATP]/(K_m^{mot}+[ATP])$ and $V_{ATP} = V_{max}^{ATP}[ATP]/(K_m^{ATP}+[ATP])$, then we obtain $d = d_{max}(K_m^{ATP}+[ATP])/(K_m^{mot}+[ATP])$, where $d_{max} = V_{max}^{mot}/V_{max}^{ATP}$. In this expression, $d$ is variable depending on the [ATP].
In fig. \ref{fig4}, we show stochastic simulations using scheme 2 with a step-size function. Rate constants were set as in the case of Fig. \ref{fig3}, except for $k_+$ = $k_-$ / $K_m^{ATP}$. We employed $d_{max}$ to be 100, 200 and 300 nm to compare with the experimental data, making $d_{max}$ five to fifteen times larger than the size of a myosin molecule. The large $d_{max}$ is expected to be due to stochastic multiple steps by myosin molecule(s), so it is actually more plausible that $d_{max}$ distributes with a large mean value (100 $\sim$ 300 nm) rather than being constant. Fig.\ref{fig4} shows a simulation with an exponential distribution and a mean of $d_{max}$. A large but constant $d_{max}$ (200 nm) can also explain the dependency of $\sigma_v^2$ on $\mu_v$, but not as well as the exponential distribution assumed here. Overall, the result shows that the model could successfully reproduce the experimental data with the most likely maximum step-size, $d_{max}$, being 200 nm. 
Thus, the characteristic the step size dependence on the [ATP] is essential to reproduce the crossover relation, confirming the "loose coupling" nature of myosin II \cite{Oosawa86}. Note that it is not essential assumption that the step is exponentially distributed, because the simulation result in which the step size is exponentially distributed but not dependent on the [ATP] condition does not show the switching behavior (Fig. \ref{fig4}, in purple). 
Here, we should again stress that our analysis does not need to estimate the number of interacting myosin molecules to obtain the displacement of an actin filament per ATP hydrolysis cycle. Although earlier studies reported the step-size, they needed to estimate the number of interacting molecules and this discrepancy lead to the one-order difference in calculated step-size \cite{Harada90,Uyeda90}.
Estimating the number of interacting molecules is extremely difficult and therefore unreliable.  Our method can overcome this difficulty, and clarify the manner of the  mechanochemical coupling. This confirms that our fluctuation analysis is a promising tool to investigate the mechanochemical coupling in molecular motors.
Also, we should note that $d_{max}$ = 200 nm is not unrealistic because Yanagida et al. and Higuchi et al. \cite{Yanagida85, Higuchi91,notice2} have already suggested the likelihood of a value much greater than that predicted from the tight coupling model \cite{Spudich94}. 

The resulting characteristics for myosin-V and myosin-II are reasonable considering their physiological functions.  Myosin-V transports cargo over a long distance by a single or a small number of motors. The tight coupling mechanism guarantees a consistent step-by-step movement over this long distance. On the other hand, myosin-II works in an ensemble like muscle to produce adaptive motions depending on the condition. The loose coupling mechanism is advantageous for such adaptation. For example, this mechanism can generate a large force with a short step at a high load but also a large step at a low load to conserve energy. Because a single myosin-II motor cannot generate steps larger than 30 nm \cite{Kitamura99}, steps as large as 200nm should be produced not by a single myosin-II motor but by the cooperation of many \cite{Julicher95}. It is possible that the active steps by some myosin molecules cause passive steps in others through the backbone connecting motors. To elucidate how the cooperative dynamics arise from the interaction of multiple motors and how this adapts to the biological environment is the focus our future work.

We thank Takao Kodama, Masahiro Ueda, Tatsuo Shibata, and Fumiko Takagi for valuable discussion,
and Peter Karagiannis for revising the manuscript.
This study was supported by "Special Coordination Funds for Promoting
Science and Technology: Yuragi Project" and Leading Project of the MEXT, Japan.


\begin{thebibliography}{99}
\bibitem{notice2}
Although their value is about three times smaller than that of ours, they use the lowest value possible for d.
\end{thebibliography}
\end{document}